\documentclass{emulateapj}
\usepackage{apjfonts}

\slugcomment{Published in ApJ, 698, 488}

\shorttitle{A MULTIWAVELENGTH STUDY OF YOUNG MASSIVE STAR-FORMING REGIONS. III.}
\shortauthors{MORALES ET AL.}

\begin{document}

\title{A Multiwavelength Study of Young Massive Star-Forming Regions. III.\\
       Mid-Infrared Emission}

\author{Esteban F. E. Morales\altaffilmark{1}, Diego Mardones and Guido Garay}
\affil{Departamento de Astronom\'ia, Universidad de Chile, Casilla 36-D, Santiago, Chile}
\email{emorales@mpifr-bonn.mpg.de}

\author{Kate J. Brooks\altaffilmark{2}}
\affil{Australia Telescope National Facility, P.O. Box 76, Epping NSW 1710, Australia}

\and

\author{Jaime E. Pineda\altaffilmark{2}}
\affil{Harvard-Smithsonian Center for Astrophysics, 60 Garden Street, MS-10, Cambridge, MA 02138}

\altaffiltext{1}{Current address: Max-Planck-Institut f\"{u}r Radioastronomie, Auf dem H\"{u}gel 69,
  53121 Bonn, Germany.}

\altaffiltext{2}{For part of this work affiliated with Departamento de Astronom\'ia, Universidad de
Chile, Casilla 36-D, Santiago, Chile}

\begin{abstract}
 
  We present mid-infrared (MIR) observations, made with the TIMMI2 camera on the ESO 3.6~m
  telescope, toward 14 young massive star-forming regions. All regions were imaged in the $N$~band,
  and nine in the $Q$~band, with an angular resolution of $ \simeq 1''$. Typically, the regions
  exhibit a single or two compact sources (with sizes in the range $0.008-0.18$~pc) plus extended
  diffuse emission. The \emph{Spitzer}--Galactic Legacy Infrared Mid-Plane Survey Extraordinaire
  images of these regions show much more extended emission than that seen by TIMMI2, and this is
  attributed to polycyclic aromatic hydrocarbon (PAH) bands. For the MIR sources associated with
  radio continuum radiation (Paper I) there is a close morphological correspondence between the two
  emissions, suggesting that the ionized gas (radio source) and hot dust (MIR source) coexist inside
  the \ion{H}{2} region. We found five MIR compact sources which are not associated with radio
  continuum emission, and are thus prime candidates for hosting young massive protostars. In
  particular, objects IRAS 14593$-$5852~II (\emph{only} detected at 17.7~\micron) and 17008$-$4040~I
  are likely to be genuine O-type protostellar objects. We also present TIMMI2 $N$-band spectra of 8
  sources, all of which are dominated by a prominent silicate absorption feature ($\simeq
  9.7$~\micron). From these data we estimate column densities in the range $(7-17)\times
  10^{22}$~cm$^{-2}$, in good agreement with those derived from the 1.2~mm data (Paper II). Seven
  sources show bright [\ion{Ne}{2}] line emission, as expected from ionized gas regions. Only
  IRAS~12383$-$6128 shows detectable PAH emission at 8.6 and 11.3~\micron.

\end{abstract}


\section{Introduction}

Massive stars are known to be born deeply embedded within dense clumps of molecular gas and dust,
with distinctive physical parameters: sizes of $\simeq 0.4$~pc, masses of $\simeq 5\times
10^3$~$M_{\sun}$, and dust temperatures of $\simeq 32$~K \citep[e.g.,][]{Faundez2004}. The formation
process of massive stars within these clumps is, however, still under discussion. Current theories
are coalescence \citep[e.g.,][]{Bonnell1998, Bonnell&Bate2002}, gravitational collapse of single
massive prestellar cores \citep[e.g.,][]{McKnee&Tan2003, Krumholz2007}, and competitive accretion
\citep[e.g.,][]{Bonnell2004}. At least up to masses corresponding to late O-type stars and early
B-type stars, the observational evidence of accretion disks and molecular outflows favors the
collapse scenario \citep[see the reviews by][]{Garay&Lizano1999, Beuther2007}. The observational
study of the early stages of high-mass stars is difficult because of three main reasons: the rapid
pre-main-sequence evolution of massive stars, confusion problems for individual studies since they
form in clusters and are located at longer distances than low-mass stars, and the large columns of
dust and gas \citep[$N($H$_2) \simeq 3\times 10^{23}$~cm$^{-2}$,][]{Garay2007} of the dense
molecular clouds that harbor them, making them completely dark at visual wavelengths. Thus, the
study of the environment around recently formed massive stars is better performed through
observations at infrared, millimeter, and radio wavelengths, where the extinction is much smaller.

To investigate the birth process and early evolution of massive stars, we are carrying out a
multiwavelength study of a sample of 18 luminous \emph{IRAS} sources in the southern hemisphere
thought to be representative of young massive star-forming regions. The goal is to understand the
physical and chemical differences between different stages of evolution. The objects were taken from
the Galaxy-wide survey of CS(2$\rightarrow$1) emission \citep{Bronfman1996} toward \emph{IRAS}
sources with far-infrared colors typical of compact \ion{H}{2} regions
\citep{Wood&Churchwell1989b}. We selected sources based primarily on the observed
CS(2$\rightarrow$1) line profiles; looking for self-absorbed profiles consistent with inward or
outward motions \citep[e.g.,][]{Mardones1998}, and/or with extended line wings, possibly indicating
the presence of bipolar outflows. In addition, the sources were required to have \emph{IRAS}
100~\micron\ fluxes greater than $10^3$~Jy and to have declination $\delta < -20\arcdeg$. The
luminosity of the \emph{IRAS} sources, computed using the \emph{IRAS} fluxes and distances derived
by L.~Bronfman (2006, private communication) are in the range $1\times10^4 - 4\times10^5$
$L_{\sun}$, implying that they contain at least one embedded massive star. The characteristics of
the radio continuum emission at $1.4-8.6$~GHz of the ionized gas and of the cold dust emission at
1.2~mm associated with the \emph{IRAS} sources in our sample are reported, respectively, in the
first and second papers of the series \citep[][hereafter Paper I and Paper II,
respectively]{Garay2006,Garay2007}.


\defcitealias{Garay2006}{Paper~I}
\defcitealias{Garay2007}{Paper~II}

Here, we report the results of mid-infrared (MIR) $N$-band and $Q$-band imaging observations, and
$N$-band spectroscopy toward a selection of sources in our original sample. The MIR continuum
emission was imaged within regions of typically $\simeq 30''\times30''$ centered on the \emph{IRAS}
sources, with an angular resolution of $\simeq 1''$. In addition, \emph{Spitzer} images at 3.6, 4.5,
5.8, 8.0, and 24~\micron\ were obtained from the public data of the galactic surveys Galactic Legacy
Infrared Mid-Plane Survey Extraordinaire (GLIMPSE) and MIPS Galactic Plane Survey (MIPSGAL).  The
primary goals of this study are to compare the MIR observations with the radio continuum images of
similar angular resolution \citepalias{Garay2006}, to determine the characteristics and physical
conditions of the warm dust surrounding recently formed massive stars, and to investigate the
stellar content within the massive and dense cores.

In Section~2, we describe the MIR observations and data reduction. Section~3 presents the images and
spectra, the overall results and computed parameters of the sample. In Section~4, we present further
discussion, and in Section~5 the conclusions of this study.


\section{Observations and Data Reduction}

The $N$- and $Q$-band images and $N$-band spectra were obtained using the TIMMI2 mid-infrared camera
\citep{Reimann2000} mounted on the ESO 3.6~m telescope at La Silla, Chile.  The camera uses a
Raytheon $320\times 240$ Si:As impurity band conduction high background array. A pixel scale of
$0.3''$ was used for $N$-band imaging observations and $0.2''$ for $Q$-band images. We performed
long-slit low-resolution spectroscopy ($70''$ north-south oriented slit, $\Delta\lambda/\lambda \sim
160$) using a pixel scale of $0.45''$ and a slit width of $1.2''$ or $3.0''$.

\begin{deluxetable}{cccrcr}
\tablecolumns{6}
\tabletypesize{\scriptsize}
\tablecaption{Summary of the TIMMI2 Observations \label{tableobs}}

\tablehead{
\colhead{IRAS source} & \colhead{Galactic name} & \colhead{Filter\tablenotemark{a}} & 
\colhead{Time} & \colhead{FOV\tablenotemark{b}} & \colhead{Noise\tablenotemark{c}}\\
\colhead{ } & \colhead{ } & \colhead{(\micron)} & \colhead{(s)} & \colhead{($['']\times['']$)} & \colhead{(mJy)}
}
\startdata
12383$-$6128  &  G301.731$+$1.104  &    8.7   &   662  &  \phd  30 $\times$ 25  &    8\\
              &                    &   11.7   &   994  &  \phd  40 $\times$ 30  &    5\\
              &                    &   17.7   &   662  &  \phd  20 $\times$ 20  &   70\\
              &                    &  8$-$13  &   767  &       3.0 $\times$ 20  &  185\\
13291$-$6249  &  G307.560$-$0.586  &    8.7   &   662  &  \phd  40 $\times$ 30  &    8\\
              &                    &   17.7   &  1027  &  \phd  20 $\times$ 15  &   35\\
              &                    &  8$-$13  &   640  &       1.2 $\times$ 15  &  187\\
14095$-$6102  &  G312.596$+$0.048  &   11.7   &   530  &  \phd  30 $\times$ 30  &    8\\
14593$-$5852  &  G319.163$-$0.419  &    8.7   &   729  &  \phd  30 $\times$ 25  &    7\\
              &                    &   17.7   &   662  &  \phd  20 $\times$ 20  &   46\\
15502$-$5302  &  G328.307$+$0.432  &   11.7   &   371  &  \phd  96 $\times$ 25  &    8\\
15520$-$5234  &  G328.808$+$0.632  &   11.7   &   795  &  \phd  40 $\times$ 30  &    5\\
              &                    &   17.7   &   580  &  \phd  30 $\times$ 20  &   41\\
              &                    &  8$-$13  &   640  &       1.2 $\times$ 20  &  119\\
16128$-$5109  &  G332.153$-$0.445  &    8.7   &    66  &  \phd  96 $\times$ 35  &   16\\
              &                    &   11.7   &   431  &  \phd  30 $\times$ 25  &    7\\
              &                    &  8$-$13  &   320  &       1.2 $\times$ 20  &  432\\
16458$-$4512  &  G340.248$-$0.373  &   11.7   &   795  &  \phd  30 $\times$ 15  &    6\\
              &                    &   17.7   &   646  &  \phd  64 $\times$ 15  &   42\\
              &                    &  8$-$13  &   640  &       1.2 $\times$ 20  &  134\\
16524$-$4300  &  G342.704$+$0.130  &   11.7   &  1325  &  \phd  30 $\times$ 15  &    4\\
17008$-$4040  &  G345.499$+$0.354  &   11.7   &   298  &  \phd  30 $\times$ 15  &    9\\
              &                    &   17.7   &   464  &  \phd  20 $\times$ 25  &   66\\
              &                    &  8$-$13  &   640  &       1.2 $\times$ 20  &  603\\
17009$-$4042  &  G345.490$+$0.311  &   11.7   &   662  &  \phd  30 $\times$ 15  &    6\\
              &                    &   17.7   &   662  &  \phd  64 $\times$ 15  &   44\\
              &                    &  8$-$13  &   639  &       3.0 $\times$ 20  &  410\\
17016$-$4124  &  G345.001$-$0.220  &   11.7   &  1275  &  \phd  40 $\times$ 30  &    4\\
              &                    &   17.7   &   397  &  \phd  20 $\times$ 15  &   50\\
              &                    &  8$-$13  &   639  &       3.0 $\times$ 20  &   81\\
17158$-$3901  &  G348.534$-$0.973  &   11.7   &  1076  &  \phd  30 $\times$ 30  &    4\\
              &                    &   17.7   &   497  &  \phd  64 $\times$ 15  &   71\\
17271$-$3439  &  G353.416$-$0.367  &   11.7   &  1656  &  \phd  40 $\times$ 30  &    3\\
\enddata

\tablenotetext{a}{8.7, 11.7 and 17.7~\micron\ refer to N1, N11.9 and Q1 images, respectively,
  whereas the $8-13$~\micron\ range corresponds to long-slit low-resolution spectroscopy.}

\tablenotetext{b}{FOV estimated depending on the observing mode. For the \emph{small-source} imaging
  mode (nodding in the east-west direction, chopping in the north-south direction): FOV = nod throw
  $\times$ chop throw. For the \emph{classic} imaging mode (both nodding and chopping in the
  north-south direction): FOV = array $x$-size ($96''$ for $N$ and $64''$ for $Q$) $\times$ nod
  throw. For spectra: FOV = slit width $\times$ nod throw.}

\tablenotetext{c}{For images, the value listed is the noise level in 1~arcsec$^2$ after flux
  calibration; for spectra, we give the average error in the spectrum that includes the whole
  emission.}

\end{deluxetable}

The observations were carried out during 2003 May 23--25. The TIMMI2 filters used were N1
($\lambda_{\mathrm{eff}} = 8.7$~\micron, $\Delta\lambda = 1.2$~\micron), N11.9
($\lambda_{\mathrm{eff}} = 11.7$~\micron, $\Delta\lambda = 1.2$~\micron), and Q1
($\lambda_{\mathrm{eff}} = 17.7$~\micron, $\Delta\lambda = 0.8$~\micron). We imaged 14 sources in at
least one filter in the $N$-band region; and nine of them in the $Q$-band region. $N$-band
spectroscopy ($8-13$~\micron) was made toward eight sources.  The standard nodding/chopping
observing technique was used to remove the strong and variable thermal background emission. For
imaging, most sources were observed in the \emph{small-source} mode (chopping in the north-south
direction, nodding in the east-west direction) using large throws ($\gtrsim 30''$), in order to
detect the extended MIR emission.  For spectroscopy, in general, smaller chop throws were selected
($\simeq 20''$).  MIR calibration standards stars --- HD~108903, HD~123139 and HD~169916 ---, were
observed in different modes (imaging and spectroscopic) every couple of hours for photometric flux
conversion and spectroscopic calibration. Table \ref{tableobs} summarizes all the observations:
Columns 1 and 2 indicate the \emph{IRAS} source identification and galactic coordinates,
respectively; Column 3 gives the TIMMI2 filter (designated by its effective wavelength
$\lambda_{\mathrm{eff}}$ or by the range $8-13$~\micron\ for the case of spectroscopy); Column 4
indicates the final on-source integration time; Column 5 indicates the field of view (FOV; east-west
amplitude $\times$ north-south amplitude for images, slit width $\times$ north-south amplitude for
spectra); and finally Column 6 lists the noise level after calibration (images: noise level in
1~arcsec$^2$, spectra: average error in the full FOV integrated spectrum).

\begin{deluxetable*}{lccrcccccc}
\tablecolumns{10}
\tabletypesize{\scriptsize}
\tablecaption{Observed Parameters\label{photTIMMI2}}

\tablehead{
\colhead{Source} & \colhead{$\alpha$} & \colhead{$\delta$} & \colhead{$\theta$ \tablenotemark{a}} & 
\colhead{$F_{\mathrm{ 8.7 \, \mu m}}^{\mathrm{ point}}$} & 
\colhead{$F_{\mathrm{ 11.7 \, \mu m}}^{\mathrm{ point}}$} & 
\colhead{$F_{\mathrm{ 17.7 \, \mu m}}^{\mathrm{ point}}$} & 
\colhead{$F_{\mathrm{ 8.7 \, \mu m}}^{\mathrm{ tot}}$} & 
\colhead{$F_{\mathrm{ 11.7 \, \mu m}}^{\mathrm{ tot}}$} & 
\colhead{$F_{\mathrm{ 17.7 \, \mu m}}^{\mathrm{ tot}}$}\\
\colhead{} & \colhead{(J2000)} & \colhead{(J2000)} & \colhead{($''$)} & 
\colhead{(mJy)} & \colhead{(mJy)} & \colhead{(mJy)} & \colhead{(mJy)} & \colhead{(mJy)} & \colhead{(mJy)}\\
}

\startdata
12383$-$6128     &   12 41 17.66  &  $-$61 44 41.0  &      2.2           &     975 $\pm$     97  &    1480 $\pm$    140  &   11300 $\pm$   2580  &    5520 $\pm$    566  &    6370 $\pm$    611  &   31800 $\pm$   7670\\
13291$-$6249     &   13 32 31.18  &  $-$63 05 18.5  &      0.7           &    3860 $\pm$    369  &        \nodata        &   19100 $\pm$   5180  &    9010 $\pm$    871  &        \nodata        &   63000 $\pm$  17100\\
14095$-$6102     &   14 13 14.22  &  $-$61 16 48.8  &  \nodata           &        \nodata        &        \nodata        &        \nodata        &        \nodata        &    3520 $\pm$    400  &        \nodata      \\
14593$-$5852 I   &   15 03 13.64  &  $-$59 04 29.9  &      3.3           &    1510 $\pm$    149  &        \nodata        &   17700 $\pm$   4880  &    4530 $\pm$    458  &        \nodata        &   45100 $\pm$  12400\\
14593$-$5852 II  &   15 03 12.60  &  $-$59 04 31.5  &3.0\tablenotemark{b}&$<81$\tablenotemark{c} &        \nodata        &    9150 $\pm$   2520  &$<285$\tablenotemark{c}&        \nodata        &   20700 $\pm$   5740\\
15502$-$5302 I   &   15 54 06.33  &  $-$53 11 40.1  &      2.0           &        \nodata        &   45300 $\pm$   4240  &        \nodata        &        \nodata        &   73600 $\pm$   6890  &        \nodata      \\
15502$-$5302 II  &   15 54 05.20  &  $-$53 11 40.4  &      2.3           &        \nodata        &    2870 $\pm$    270  &        \nodata        &        \nodata        &    4510 $\pm$    425  &        \nodata      \\
15520$-$5234     &   15 55 48.28  &  $-$52 43 06.7  &      1.1           &        \nodata        &    4780 $\pm$    505  &   19300 $\pm$   6610  &        \nodata        &   11400 $\pm$   1210  &   73000 $\pm$  25000\\
16128$-$5109     &   16 16 40.22  &  $-$51 17 11.8  &  \nodata           &    2240 $\pm$    242  &    6210 $\pm$    661  &        \nodata        &   29400 $\pm$   3680  &  101000 $\pm$  10700  &        \nodata      \\
16458$-$4512     &   16 49 30.02  &  $-$45 17 44.4  &      1.2           &        \nodata        &    3010 $\pm$    294  &   13100 $\pm$   3180  &        \nodata        &    5860 $\pm$    577  &   28400 $\pm$   6960\\
16524$-$4300     &   16 56 03.47  &  $-$43 04 41.6  &  \nodata           &        \nodata        &     348 $\pm$     40  &        \nodata        &        \nodata        &    1990 $\pm$    234  &        \nodata      \\
16547$-$4247     &   16 58 16.55  &  $-$42 52 04.2  &  \nodata           &        \nodata        &        \nodata        &        \nodata        &        \nodata        &280 $\pm$ 28\tablenotemark{d}&  \nodata      \\
17008$-$4040 I   &   17 04 22.82  &  $-$40 44 22.6  &      0.9           &        \nodata        &   15000 $\pm$   1630  &   62800 $\pm$  14200  &        \nodata        &   18300 $\pm$   1990  &   80200 $\pm$  18100\\
17008$-$4040 II  &   17 04 23.48  &  $-$40 44 35.7  &      1.4           &        \nodata        &     861 $\pm$    100  &    5240 $\pm$   1210  &        \nodata        &    1010 $\pm$    119  &    5360 $\pm$   1280\\
17009$-$4042     &   17 04 28.02  &  $-$40 46 24.8  &      1.6           &        \nodata        &    9570 $\pm$   1060  &   45700 $\pm$  10400  &        \nodata        &   22600 $\pm$   2510  &  141000 $\pm$  32200\\
17016$-$4124 I   &   17 05 11.18  &  $-$41 29 06.6  &      1.1           &        \nodata        &     951 $\pm$    102  &    7620 $\pm$   2370  &        \nodata        &    1580 $\pm$    174  &   14900 $\pm$   4660\\
17016$-$4124 II  &   17 05 10.98  &  $-$41 29 14.1  &      1.1           &        \nodata        &     857 $\pm$     92  &    7100 $\pm$   2210  &        \nodata        &    1060 $\pm$    116  &    7880 $\pm$   2490\\
17158$-$3901     &   17 19 15.47  &  $-$39 04 32.5  &      1.2           &        \nodata        &    1460 $\pm$    155  &    6240 $\pm$   1880  &        \nodata        &    8780 $\pm$    936  &   25400 $\pm$   7770\\
17271$-$3439     &   17 30 28.55  &  $-$34 41 48.8  &      0.7           &        \nodata        &     838 $\pm$     89  &        \nodata        &        \nodata        &   10900 $\pm$   1170  &        \nodata      \\
\enddata

\tablecomments{Units of right ascension are hours, minutes, and seconds, and units of declination
  are degrees, arcminutes, and arcseconds. The position is from the brightest component when the
  source is multiple; astrometry was performed using the \emph{Spitzer}--IRAC
  images. $F^{\mathrm{point}}$ is the flux density calculated on compact sources using $2.0''$
  radius aperture photometry. $F^{\mathrm{tot}}$ is the flux density calculated on a region
  enclosing the total emission from each source.  1$\sigma$ errors are given, taking into account
  statistical and calibration uncertainties.}

\tablenotetext{a}{The angular size $\theta$ is the deconvolved FWHM, computed as $\theta =
  \sqrt{\mathrm{FWHM}^2 - \mathrm{FWHM_{std}}^2}$, where FWHM is determined on the $N$-band image
  (11.7~\micron\ when available, 8.7~\micron\ otherwise) for the compact source, and
  FWHM$_\mathrm{std}$ is the averaged value of the standard stars in the corresponding filter.}

\tablenotetext{b}{Angular size computed at 17.7~\micron.}

\tablenotetext{c}{3$\sigma$ upper limit.}

\tablenotetext{d}{From \citet{Brooks2003}, assuming a flux uncertainty of 10\%.}

\end{deluxetable*}

The images were reduced using our own IDL routines based on scripts originally developed by
M. Marengo (2004, private communication). Basically, the chopping and nodding pairs corresponding to
a single chop-nod cycle were co-added in an individual frame, and then each useful portion in that
image was bari-centered before the combination of all the frames.  This prevented us from losing
angular resolution due to little offsets between different chop-nod cycles.  The final image for
each source was produced co-adding all the observations weighted by the integration times. The flux
calibration was made using the tabulated fluxes for the standard stars given on the instrument Web
page \citep[based on preliminary versions of the spectral models by][]{Cohen1999}. Finally, the
images were Gaussian smoothed to a resolution of $\simeq$ $1.8''$, in order to enhance the
sensitivity to diffuse extended emission and to match the angular resolution of the
\emph{Spitzer}--GLIMPSE images (see below), allowing us to compare both sets of data in a consistent
way.

For spectra reduction we used the IDL scripts provided by \citet{Siebenmorgen2004}, slightly
modified in order to include the extended emission in the spectra extraction.  This method is based
on the optimal extraction described by \citet{Horne1986}, and consists of summing along the spatial
dimension after weighting by a source profile, which is computed by collapsing the image along the
dispersion direction.  Wavelength calibration was done mainly using the tables given in the Web
page. The same procedure was applied to the target and the corresponding standard star, whose
theoretical spectrum \citep[models by][]{Cohen1999} was used to calibrate the final target spectrum.

We complemented our observations with MIR data obtained by the \emph{Spitzer Space Telescope}
\citep{Werner2004}. Images at 3.6, 4.5, 5.8, and 8.0~\micron\ were downloaded from the public data
of the GLIMPSE Legacy Program \citep{Benjamin2003}, that surveyed the plane in the four bands of the
Infrared Array Camera \citep[IRAC;][]{Fazio2004}. We also obtained images at 24~\micron\ from the
recently delivered mosaics of the MIPSGAL Legacy Program \citep{Carey2009}, carried out using the
Multiband Imaging Photometer for \emph{Spitzer} \citep[MIPS;][]{Rieke2004}. The 18 sources of our
multiwavelength survey are within the $(l,b)$ range covered by GLIMPSE/GLIMPSE~II and MIPSGAL,
except IRAS~12383$-$6128 which has only data at 4.5 and 8.0~\micron. The angular resolutions of the
used \emph{Spitzer} filters, calculated from the point-spread function (PSF) archives given in the
telescope Web page, are $1.81''$, $1.76''$, $1.84''$, $2.14''$, and $5.81''$ at 3.6, 4.5, 5.8, 8.0,
and 24~\micron, respectively.

\begin{figure*}[!htp]

\centering
\includegraphics[height= 23cm]{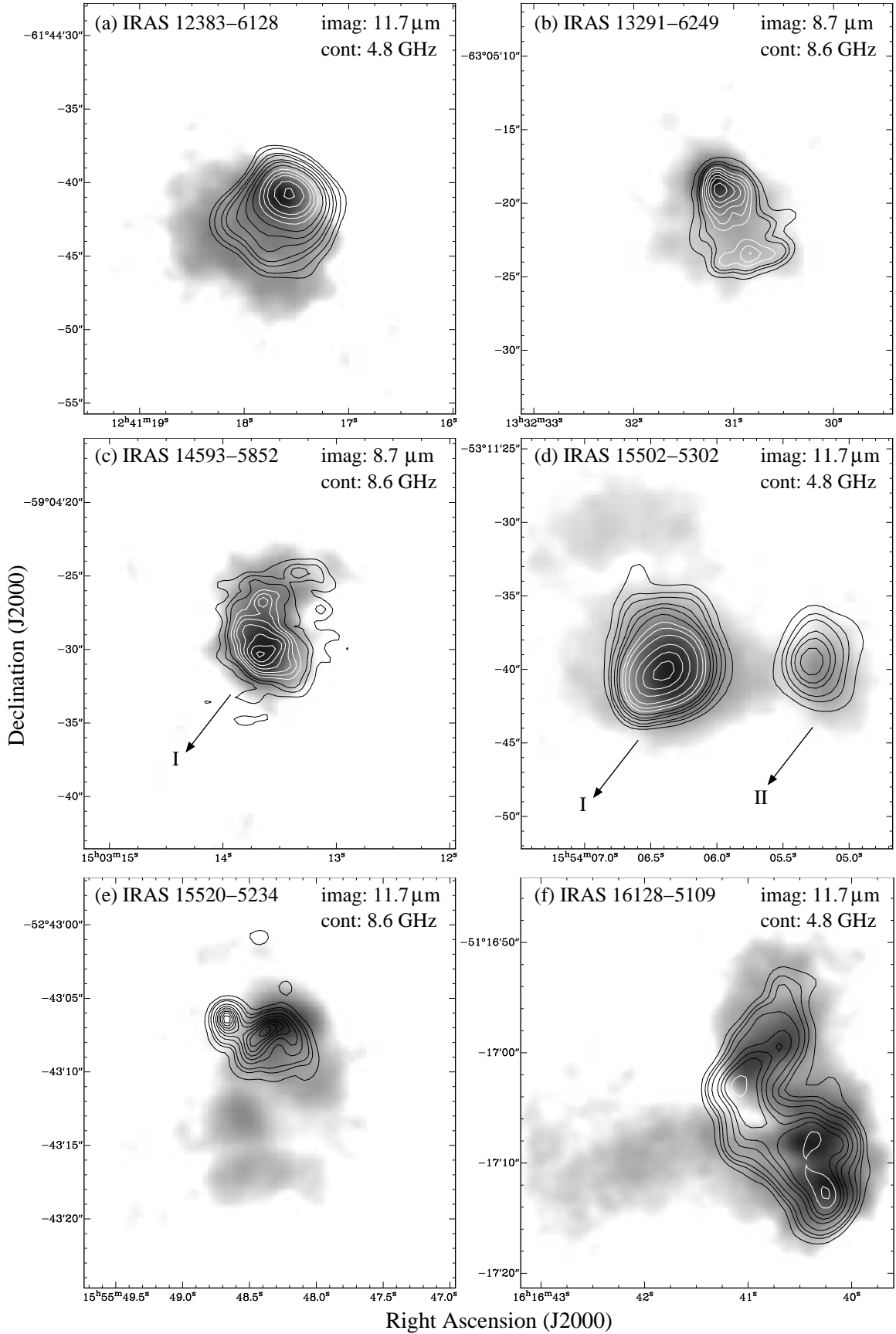}

\caption{TIMMI2 $N$-band gray-scale images, overlaid with radio continuum ATCA contours. Source
  names are showed in the upper left corner, whereas the upper right corner indicates the specific
  TIMMI2 filter used (11.7 or 8.7~\micron) and the frequency of the radio contours (4.8 or 8.6
  GHz). The images are displayed in logarithmic scale, over the full flux range. Contour levels are
  basically the same than those used in \citetalias{Garay2006}.}

\label{TIMMI2images1}
\end{figure*}

\begin{figure*}[!htp]

\centering
\includegraphics[height= 23cm]{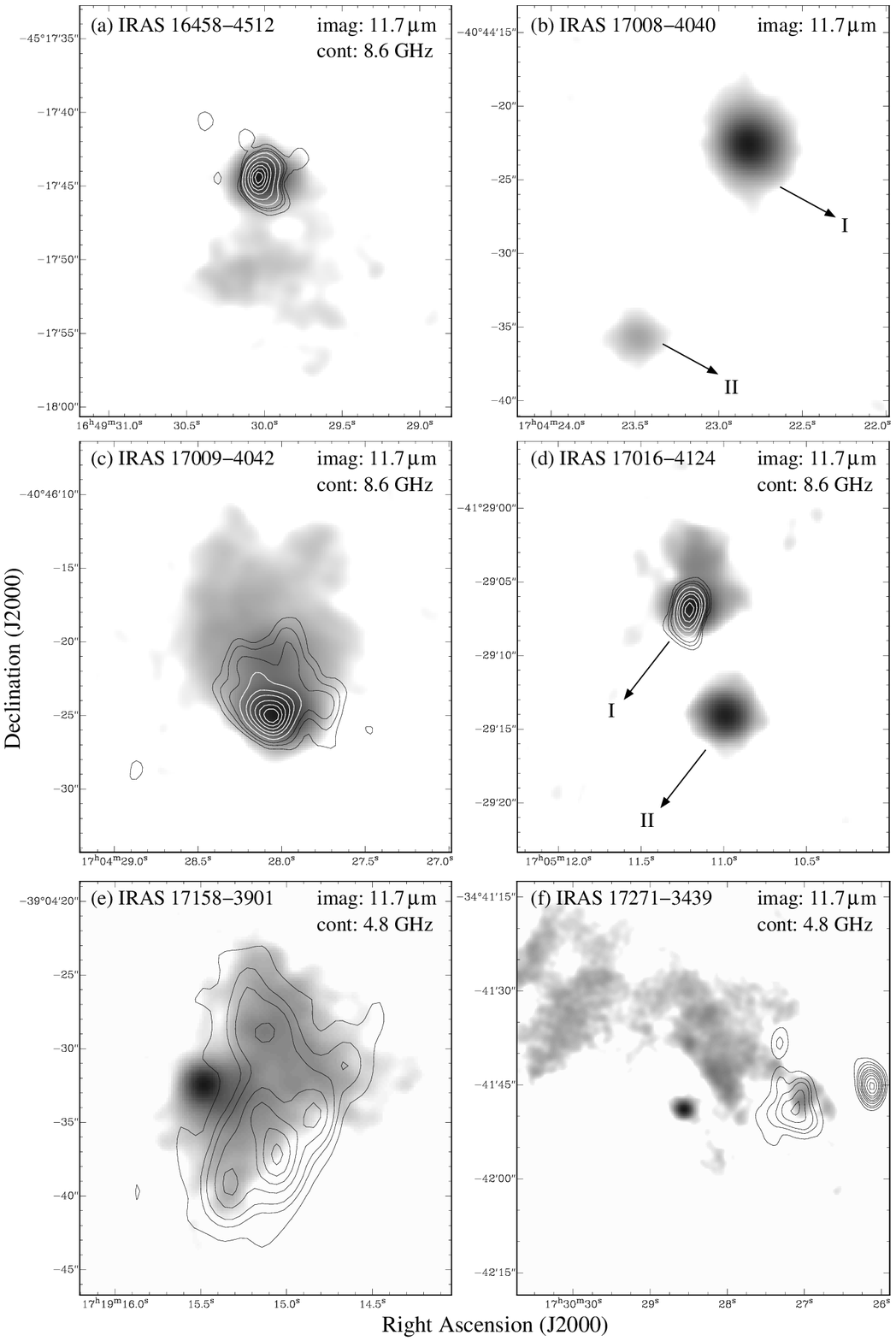}

\caption{Same as in Figure \ref{TIMMI2images1}. (\emph{b}) For IRAS~17008$-$4040, no radio continuum
  emission was detected toward objects I and II up to a level of 1.5 mJy at 4.8 GHz.}
\label{TIMMI2images2}
\end{figure*}

\section{Results}

\subsection{TIMMI2 Images}\label{secTIMMI2im}

\begin{figure*}[!htp]
\centering
\includegraphics[height= 8cm]{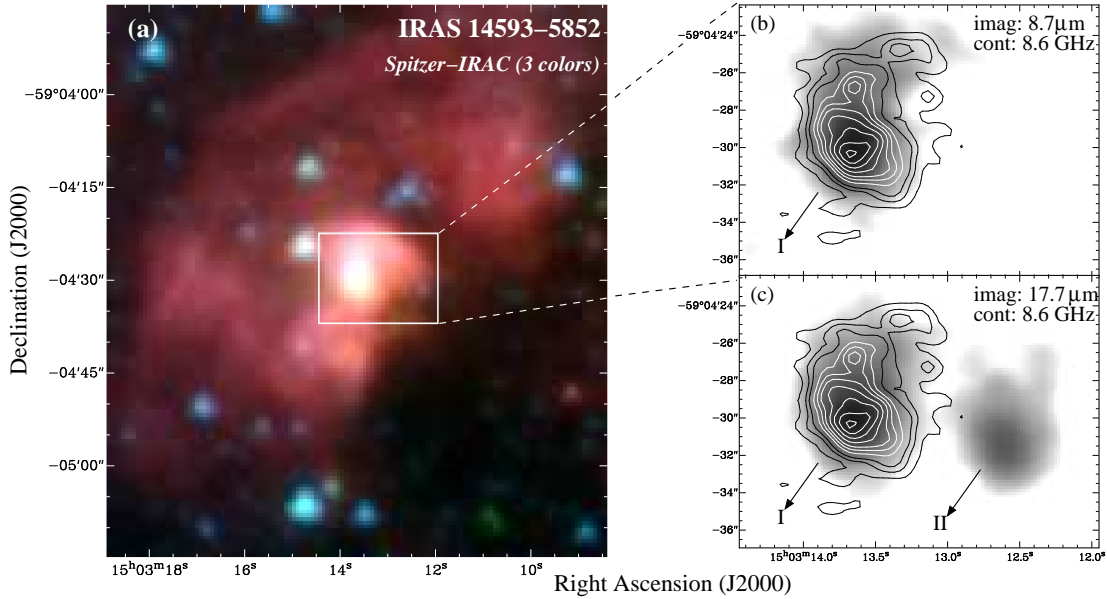}

\caption{IRAS~14593$-$5852. (\emph{a}) \emph{Spitzer}--IRAC image made using 3.6 (blue), 4.5
  (green), and 8.0~\micron\ (red) bands. (\emph{b}) TIMMI2 8.7~\micron\ gray-scale image, overlaid
  with ATCA 8.6~GHz contours. (\emph{c}) TIMMI2 17.7~\micron\ gray-scale image, overlaid with ATCA
  8.6~GHz contours.}
\label{14593}
\end{figure*}

We detected MIR emission toward the observed regions in all the observed
filters. Figures~\ref{TIMMI2images1} and \ref{TIMMI2images2} show gray-scale images of the smoothed
emission in the $N$-band (8.7~\micron\ or 11.7~\micron) toward 12 regions. We do not show images of
the emission toward IRAS~14095$-$6102 and 16524$-$4300, which have a low signal-to-noise ratio
(S/N). In seven regions, we detected a single MIR source. Most of these have a cometary-like
morphology, exhibiting a bright compact component at the head and a weak trailing extended
emission. In three regions (IRAS~15502$-$5302, 17016$-$4124, and 17008$-$4040) we detected the
presence of two distinct MIR objects, and in two regions (IRAS~16128$-$5109 and 17271$-$3439) the
MIR emission has a complex morphology. The morphologies of the sources detected in both the $N$ and
$Q$ bands are very similar. The only exception is in the IRAS~14593$-$5852 region, where we detected
a source at 17.7~\micron\ (14593$-$5852~II) which is not seen in the $N$ filter (see
Figure~\ref{14593}).

Among the compact components, no one is strictly pointlike: whereas the average FWHM size of the
standard stars is $0.9''$ in the 8.7~\micron\ filter, $1.2''$ at 11.7~\micron, and $1.3''$ at
17.7~\micron, the compact components have FWHM sizes $\gtrsim 1.4''$ and $\gtrsim 1.5''$ at
11.7~\micron\ and 17.7~\micron, respectively. However, we do not discard the presence of point
sources in a few cases, due to the uncertainties in the FWHM determination, which was made using a
radial profile fitting.

According to \citet{Saviane&Doublier2005}, the pointing accuracy of the ESO~3.6~m telescope is
$\simeq 5''$ which prevents a direct comparison of the MIR and radio continuum spatial
distributions. We used the \emph{Spitzer}--IRAC data to determine accurate astrometry of our images.
The bright compact part of the smoothed TIMMI2 image was fitted to the corresponding peak in the
longest wavelength IRAC image available (usually the 8.0 \micron\ image). In most cases, the
morphology of the emission exhibited in the IRAC and our images are very similar, allowing us to
match the positions without any confusion. The offsets between the original TIMMI2 coordinates and
the new adjusted positions are typically $\simeq 11''$. Since the GLIMPSE point-source accuracy is
$\simeq 0.3''$ \citep{Meade2005}, we estimate that our final TIMMI2 coordinates have an error
$\lesssim 1''$, produced by centering uncertainties (e.g., corresponding regions not compact enough
or peaks not well defined).

Figures \ref{TIMMI2images1} and \ref{TIMMI2images2} also show contour maps of the radio continuum
emission \citepalias[][4.8 or 8.6~GHz]{Garay2006} overlaid on the $N$-band images. These figures
show that in most sources the spatial distribution of the radio and MIR emissions are highly
correlated, both in extended (IRAS~12383$-$6128, 13291$-$6249, 14593$-$5852, 15502$-$5302,
16128$-$5109, and 17009$-$4042) and compact structures (the same sources, together with
IRAS~15520$-$5234, 16458$-$4512, and 17016$-$4124~I). This suggests an intimate association between
the ionized gas and the warm dust within these regions (see further discussion in
Section~\ref{secMIRHII}). Although in five (of 10) compact sources the corresponding radio and MIR
peaks are slightly shifted, the displacement never reaches more than $\simeq 1''$ and therefore does
not affect the global morphological match. Since these shifts are small, they probably can be
explained uniquely by positional uncertainties of the data. The exceptional cases in which an MIR
source is not associated with radio continuum emission will be discussed in Section~\ref{secHMPO}.

Table \ref{photTIMMI2} lists the source properties computed directly from the TIMMI2 images
(positions, angular sizes, and flux densities). For each source, the given position corresponds to
the centroid of the compact component; the brightest compact object is taken as a reference when the
region is complex. The angular size was calculated as $\theta = \sqrt{\mathrm{FWHM}^2 -
  \mathrm{FWHM_{std}}^2}$, where FWHM is the full width half-maximum of the compact component and
FWHM$_\mathrm{std}$ is the averaged value of the observed standard stars, both calculated by fitting
the radial profile on the $N$-band image (11.7~\micron\ when available, 8.7~\micron\ otherwise). We
computed two flux densities: a ``point-source'' flux density $F^{\mathrm{point}}$, calculated on
compact sources using a $2.0''$ radius aperture photometry (diameter of about twice the smoothed
resolution of $1.8''$), and a ``total'' flux density $F^{\mathrm{tot}}$, calculated on a region
(circular in almost all cases) enclosing the whole emission from each source. Photometric error was
computed adding by quadrature the statistical noise of the image and the flux calibration error,
which is mainly determined by the atmospheric fluctuations.  We estimated this contribution as the
standard deviation of the calibration factor along the three nights of observation. The resulting
total photometric error is $\simeq 10\%$ for the 8.7 and 11.7~\micron\ filters and $22\%-34\%$ for
the 17.7~\micron\ filter, which is affected by the presence of many atmospheric features within the
filter wavelength range.

Within the 14 regions observed with TIMMI2, we can identify at least 15 simple bright compact
sources with $N$-band deconvolved FWHM sizes in the range $0.7''-3.3''$ ($\theta$ in Table
\ref{photTIMMI2}). Using the kinematical distances, the physical diameters of these objects are in
the range $0.008-0.18$~pc, with an average value of $0.03$~pc. The physical sizes of the compact MIR
components associated with radio continuum sources are in good agreement with each other, when
measured at a comparable angular resolution (at 8.6~GHz). This supports the idea that the radio and
MIR emission trace similar structures.

\begin{figure*}[!htp]

\centering
\includegraphics[height= 8cm]{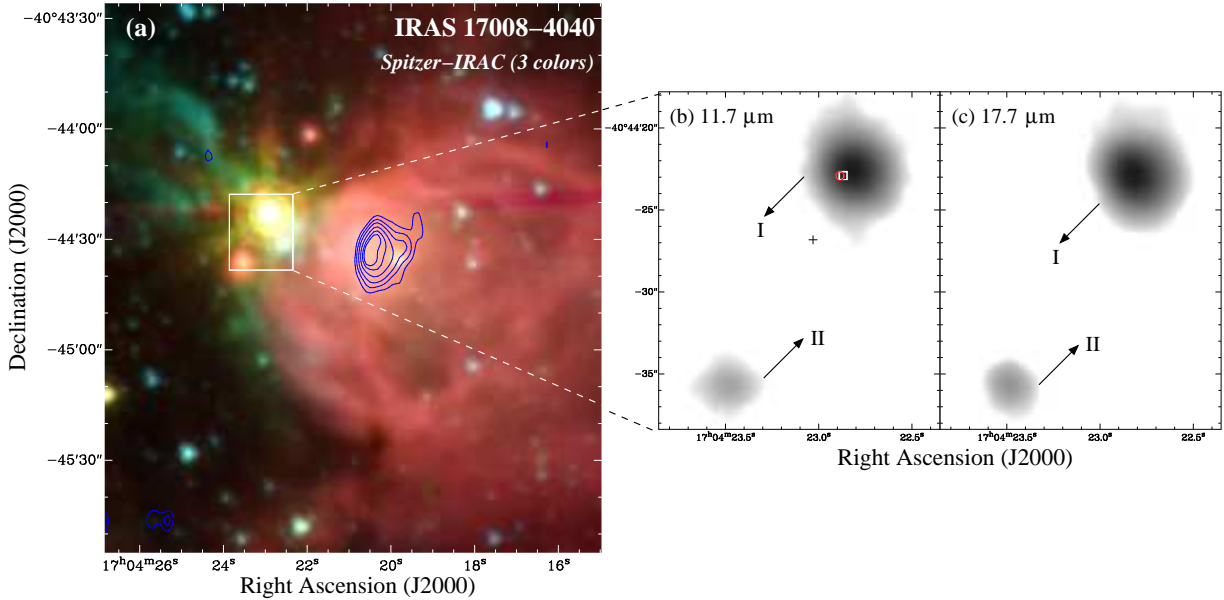}

\caption{IRAS~17008$-$4040. (\emph{a}) \emph{Spitzer}--IRAC image made using 3.6 (blue), 4.5
  (green), and 8.0~\micron\ (red) bands. Also shown are ATCA contours (in blue) at
  2.5~GHz. (\emph{b}) TIMMI2 11.7~\micron\ gray-scale image, with the locations of maser sites
  indicated by small symbols: OH main-line by a white square \citep{Caswell1998}, 6.7-GHz methanol
  by a red circle \citep{Walsh1998}, and water vapor by a black cross
  \citep{Forster&Caswell1989}. (\emph{c}) TIMMI2 17.7~\micron\ gray-scale image.}
\label{17008}
\end{figure*}

\begin{deluxetable}{lrrrrr}
  \tablecolumns{6} \tabletypesize{\footnotesize} \tablecaption{\emph{Spitzer}
    Fluxes\label{photIRAC}} 
\tablehead{
\colhead{Source} & 
\colhead{$F_{\mathrm{ 3.6 \, \mu m}}^{\mathrm{ tot}}$} & 
\colhead{$F_{\mathrm{ 4.5 \, \mu m}}^{\mathrm{ tot}}$} & 
\colhead{$F_{\mathrm{ 5.8 \, \mu m}}^{\mathrm{ tot}}$} & 
\colhead{$F_{\mathrm{ 8.0 \, \mu m}}^{\mathrm{ tot}}$} &
\colhead{$F_{\mathrm{ 24 \, \mu m}}$} \\
\colhead{ } & \colhead{(mJy)} & \colhead{(mJy)} & \colhead{(mJy)} & \colhead{(mJy)} & \colhead{(mJy)}
}
\startdata
12383$-$6128            &  \nodata\tablenotemark{a}  &      460  &  \nodata\tablenotemark{a}  &     7980  &                        \nodata\tablenotemark{a}\\
13291$-$6249            &     1140  &$>1990$\tablenotemark{b}  &     6290  &$>8370$\tablenotemark{b}  &                           $>32400$\tablenotemark{c}\\
14095$-$6102            &      158  &      219  &     1770  &     4410  &                                                            69300\tablenotemark{d}\\
14593$-$5852 I          &      255  &      411  &     1670  &     4660  &                                                        $>32100$\tablenotemark{ce}\\
14593$-$5852 II         &$<33$\tablenotemark{f}&$<44$\tablenotemark{f}&$<347$\tablenotemark{f}&$<950$\tablenotemark{f}  &                           \nodata\\
15394$-$5358            &      878  &5050\tablenotemark{d}  &$>7710$\tablenotemark{b}  & 9840\tablenotemark{d}  &                    46900\tablenotemark{d}\\
15502$-$5302 I          &     1160  &$>2380$\tablenotemark{b}  &    11000  &$>13300$\tablenotemark{b}  &                         $>27800$\tablenotemark{ce}\\
15502$-$5302 II         &      111  &      174  &      802  &     2090  &                                                                           \nodata\\
15520$-$5234            &      315  &      936  &     2760  &     6560  &                                                           556000\tablenotemark{d}\\
15596$-$5301            &      128  &      352  &     1040  &     1870  &                                                            44900\tablenotemark{d}\\
16128$-$5109            &     2860  &     3950  &    17700  &    53300  &                                                         $>12300$\tablenotemark{c}\\
16272$-$4837            &       87  &      317  &      432  &      389  &                                                                              6040\\
16458$-$4512            &      188  &      423  &     1890  &     4530  &                                                            76000\tablenotemark{d}\\
16524$-$4300            &      366  &      955  &     2940  &     5410  &                                                            65800\tablenotemark{d}\\
16547$-$4247            &      119  &      564  &      988  &      913  &                                                            49900\tablenotemark{d}\\
17008$-$4040 I          &      557  &9640\tablenotemark{d}  &23700\tablenotemark{d}  &45100\tablenotemark{d}  &                  $>49500$\tablenotemark{ce}\\
17008$-$4040 II         &       27  &       53  &      359  &      883  &                                                                           \nodata\\
17009$-$4042            &      470  &     1560  &     5920  &$>12800$\tablenotemark{b}  &                                         $>63800$\tablenotemark{c}\\
17016$-$4124 I          &       71  &      442  &     1010  &     2040  &                                                          195000\tablenotemark{de}\\
17016$-$4124 II         &       25  &       48  &      242  &      677  &                                                                           \nodata\\
17158$-$3901            &      488  &      528  &     3000  &     8350  &                                                         $>54300$\tablenotemark{c}\\
17271$-$3439            &     1020  &     1430  &     8440  &    24600  &                                                          $>5130$\tablenotemark{c}\\
\enddata

\tablecomments{$F^{\mathrm{tot}}$ is the flux density calculated on the same region as for TIMMI2
  photometry. Uncertainties are estimated to be $\simeq 20\%$ for the nonsaturated IRAC bands.}

\tablenotetext{a}{Out of GLIMPSE/MIPSGAL coverage.} 

\tablenotetext{b}{Saturated; lower limit obtained integrating the emission in the TIMMI2-defined region.}

\tablenotetext{c}{Saturated; lower limit obtained using $35''$ radius aperture photometry.}

\tablenotetext{d}{Saturated; recovered flux using the \emph{imageworks} PSF fitting tool (see the
  text).}

\tablenotetext{e}{Double sources not resolved. Flux includes components I and II.}

\tablenotetext{f}{Upper limit obtained integrating the diffuse emission in the TIMMI2-defined
    region.}

\end{deluxetable}

\subsection{Spitzer Images}\label{secGLIMPSE}

\emph{Spitzer} and $Q$-band images of all sources are shown in a dedicated Web page
\footnote{\vspace{-0.5cm}
  \url{http://www.mpifr-bonn.mpg.de/staff/emorales/mir\_individual/mir\_individual.html}}.  In this
Web page we further discuss the characteristics of the MIR emission toward each of the individual
regions of the survey and summarize what is already known about them in the literature. In general,
the IRAC images show similar structures to those seen in the TIMMI2 images, although the former are
much more sensitive to diffuse extended emission, at 5.8 and 8.0~\micron. To illustrate this, we
show in Figures~\ref{14593} and \ref{17008} two examples of three-color images made combining the
individual 3.6, 4.5, and 8.0~\micron\ IRAC bands, together with the corresponding $N$ and $Q$-band
images from TIMMI2. These two regions are those containing the most promising high-mass protostellar
object (HMPO) candidates (see Section~\ref{secHMPO}).  The diffuse emission seen in the 5.8 and 8.0
\micron\ IRAC filters, not detected with TIMMI2, is not associated with radio continuum emission
from ionized gas and probably contains a considerable contribution from polycyclic aromatic
hydrocarbon (PAH) emission bands at 6.2, 7.7, and 8.6~\micron\ (see Section~\ref{secPAH}). In
addition, due to its higher sensitivity and shorter wavelengths available, the IRAC images reveal
that in most cases the TIMMI2 source is the brightest one of a cluster of MIR objects.

Some of the three-color images show the presence of IR features that appear greenish (the so-called
green fuzzies), most of which are extended and not associated with radio continuum emission. This
type of emission is found toward IRAS~15394$-$5358, 15520$-$5234, 15596$-$5301, 16272$-$4837,
16547$-$4247, 17008$-$4040 (Figure~\ref{17008}), and 17016$-$4124. \citet{Beuther2005} and
\citet{Rathborne2005} first suggested that the bright and extended IRAC 4.5~\micron\ emission might
trace shocked gas by outflowing material ramming in the ambient interstellar medium.  Indeed,
numerical simulations predict that the contribution of H$_2$ shock-excited line emission in the
4.5~\micron\ band is an order of magnitude brighter than in the other IRAC bands
\citep{Smith&Rosen2005}.  Observational support for this interpretation has been provided by recent
\emph{Spitzer} surveys \citep[e.g.,][]{Noriega2004, Smith2006}, which show that molecular outflows
are particularly strong in the 4.5~\micron\ band.  We also found that most of the sources which
exhibit green fuzzies (all but IRAS 15596$-$5301 and 17008$-$4040) are embedded in large-scale
8.0~\micron\ extinction features known as infrared dark clouds (IRDCs), which host the earliest
stages of high-mass star formation \citep[see][and references therein]{Rathborne2005}.

All the sources of our multiwavelength survey covered by MIPSGAL present very bright emission at
24~\micron, quite saturated at the core of the PSF (with the exception of IRAS~16272$-$4837); in
particular, the TIMMI2 objects lie just within these saturated cores. In some regions, interesting
large-scale structures could be seen at 24~\micron, hinting the presence of warm dust probably
excited by our target sources beyond the FOV studied with TIMMI2. We tried to overcome the
saturation and rescue the lost MIPS fluxes using the publicly available \emph{imageworks} code
\footnote{\url{http://spider.ipac.caltech.edu/staff/jarrett/irac/index-1.html}} written by
T. Jarrett, which recovers the saturated pixels by applying a PSF model that is fit to the
nonsaturated wings of the source. However, since most of our sources are embedded, the PSF wings are
often contaminated by diffuse emission and the recovered flux was overestimated by the code. We
therefore only consider valid those fluxes which are consistent with our spectral energy
distribution (SED) models, as described in Section~\ref{secSED}; they are listed in the last column
of Table \ref{photIRAC}. For the remaining sources, we give lower limit fluxes, obtained by
integrating the emission of the PSF wings within an aperture of $35''$ and applying the aperture
correction of \citet{Engelbracht2007}.

Table \ref{photIRAC} gives also the results of the photometry obtained from the IRAC images. In
order to compare properly with TIMMI2 data, we integrated the emission across the same regions as
for the TIMMI2 photometry, even though the emission in the IRAC images is usually more extended.
This resulted, however, in a good correlation between \emph{MSX} 8.3~\micron\ and IRAC 8.0~\micron\
fluxes (see Section~\ref{secSED}), indicating that the regions selected for photometry were
appropriate.  For the sources not observed with TIMMI2, we chose similar apertures ($\simeq 6''-9''$
radius) enclosing the brightest source(s). According to \citet{Meade2005}, magnitude uncertainties
are estimated to be less than 0.2 for the GLIMPSE Catalog, corresponding to a flux error of
$\lesssim 18\%$.  Considering that our sources are not pointlike, we estimated our photometric
uncertainty as 20\%.  For the saturated sources, we proceeded as for MIPS photometry, applying the
\emph{imageworks} code and considering the recovered flux only if it is consistent with the SED
model; otherwise, we give a lower limit flux integrated within the TIMMI2-defined region.

\subsection{TIMMI2 Spectra}\label{secTIMMI2spe}

\begin{figure*}[!htp]

\centering
\includegraphics[height= 23cm]{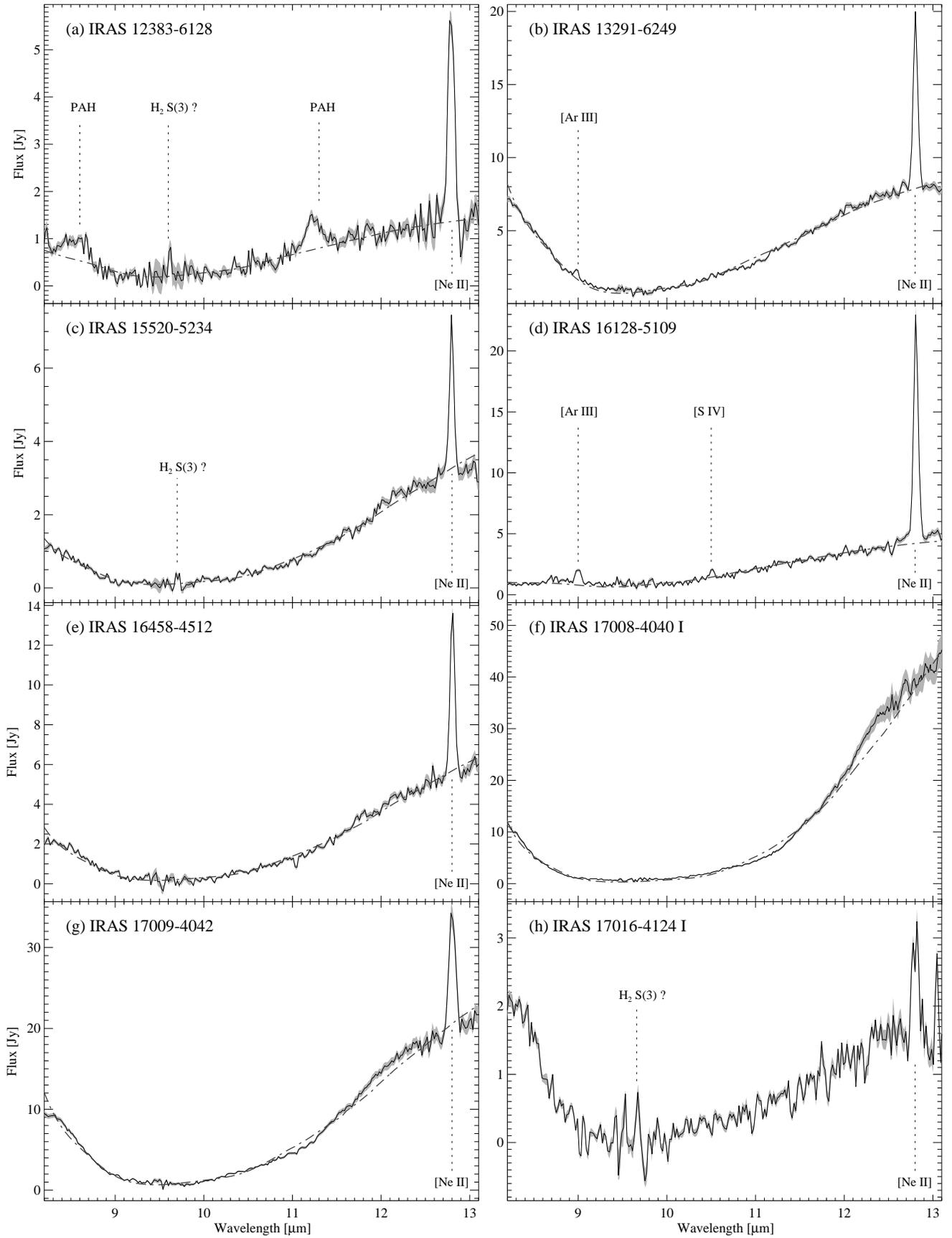}

\caption{TIMMI2 8$-$13~\micron\ spectra of the central parts of the sources. The statistical noise
  is shown as a filled region around the main curve. The spectral features found are labeled and
  marked with vertical dashed lines. We also show here the best fit model of the silicate absorption
  band (dashed curve) according to the procedure explained in Section~\ref{secTIMMI2spe}.}

\label{TIMMI2spectra}
\end{figure*}

Figure \ref{TIMMI2spectra} presents the $8-13$~\micron\ spectra of the eight observed sources,
extracted within small angular apertures of $\simeq 4'' - 8''$ around the central positions of the
spectral images. The north-south oriented slit was centered in the brightest part of each source.
The spectra integrated in the whole emission regions, not shown here, exhibit practically the same
shape and features, but logically with a different flux scale. We truncated the spectra at
8.2~\micron\ due to calibration uncertainties at the lower wavelength edge. Error bars are plotted
as filled regions around the main curve; they only represent the statistical noise derived from the
original spectral image and do not include other causes of error.

All the spectra exhibit a broad --- and in most cases prominent --- silicate band seen in
absorption, at $\simeq 9.7$~\micron.  Seven sources (all but IRAS~17008$-$4040~I,
Figure~\ref{TIMMI2spectra}(f)) present [\ion{Ne}{2}] line emission at $\simeq 12.8$~\micron, which
is spatially extended.  This is consistent with the association of these sources with radio
continuum emission of ionized gas.  Column 2 of Table \ref{tableSpectra} gives the [\ion{Ne}{2}]
line flux measured on the integrated spectrum of each source; the statistical uncertainties derived
directly from the spectrum are $\lesssim 2\%$. Emission in other two atomic fine-structure lines,
[\ion{Ar}{3}] at 9.0~\micron\ and [\ion{S}{4}] at 10.5~\micron, was observed toward IRAS
16128$-$5109 (Figure~\ref{TIMMI2spectra}(d), but this is more evident in the integrated
spectrum). The [\ion{Ar}{3}] is also present in IRAS~13291$-$6249 (Figure~\ref{TIMMI2spectra}(b)).
The PAH features at 11.3~\micron\ and 8.6~\micron\ were detected in only one region
(IRAS~12383$-$6128, Figure~\ref{TIMMI2spectra}(a)). The faint emission feature at $\simeq
9.7$~\micron\ in IRAS~13291$-$6249, 15520$-$5234, and 17016$-$4124~I
(Figures~\ref{TIMMI2spectra}(a), (c), and (h), respectively) could correspond to the H$_2$ 0--0
$S(3)$ line, but it is located in a wavelength range highly contaminated by telluric ozone features,
which make noisier this part of the spectrum (note that this is in general not properly represented
by the statistical noise shown in the spectra). We therefore label this line with a question mark in
Figure~\ref{TIMMI2spectra}. If present, the H$_2$ 0--0 $S(3)$ line could be tracing a
photodissociation region or the existence of shock events, since it can be either radiatively or
collisionally excited \citep[see, e.g.,][]{Morris2004}.

We fitted the continuum part of the TIMMI2 spectrum (i.e., removing first the line features) using a
simple model \citep{Pascucci2004} which assumes that the MIR continuum flux arises from a small
region of hot dust, whose emission can be described by a gray body with temperature
$T_{\mathrm{d}}$, column density $N(\mathrm{H}_2)_{\mathrm{hot}}$, and solid angle
$\Omega_{\mathrm{d}}$, surrounded by a larger cloud of cold dust with column density
$N(\mathrm{H}_2)_{\mathrm{cold}}$.  Under these assumptions, the flux density as a function of
frequency is given by
\begin{equation}
F_\nu = \Omega_{\mathrm{d}} B_\nu(T_{\mathrm{d}})(1 - e^{-\sigma_\nu N(\mathrm{H}_2)_{\mathrm{hot}}})
        e^{-\sigma_{\nu} N(\mathrm{H}_2)_{\mathrm{cold}}/2}\label{eq_flux_sil}~~,
\end{equation}
where
\begin{eqnarray}
\sigma_{\nu} &=& R_{\mathrm{dg}}\mu m_{\mathrm{H}}\, 
\kappa_{\nu} \label{eq_sigma_nu}~~.
\end{eqnarray}

\noindent
$R_{\mathrm{dg}}$ is the dust-to-gas mass ratio, $\mu$ is the mean molecular weight per hydrogen
molecule, $m_{\mathrm{H}}$ is the hydrogen mass, and $\kappa_{\nu}$ is the absorption opacity per
mass of dust, whose tabulated values were taken from the \citet{Weingartner&Draine2001} dust model,
with $R_V = 5.5$. \citet{Chapman2009} found that this model is more consistent with the MIR
extinction law of high-density regions than the \citet{Weingartner&Draine2001} model with $R_V =
3.1$, because it includes larger dust grains and therefore accounts for possible grain growth. The
fitting procedure uses equation (\ref{eq_flux_sil}) with $\Omega_{\mathrm{d}}$, $T_{\mathrm{d}}$,
and $N(\mathrm{H}_{2})_{\mathrm{cold}}$ as free parameters.  We constrain the parameter
$N(\mathrm{H}_{2})_{\mathrm{hot}}$ by imposing the condition that
\begin{equation}
N(\mathrm{H}_{2})_{\mathrm{hot}} = \sqrt{\frac{\Omega_{\mathrm{d}}}
{\Omega_{\mathrm{c}}}}N(\mathrm{H}_{2})_{\mathrm{cold}} \label{eq_ncte_sil}~~,
\end{equation}
where $\Omega_{\mathrm{c}}$ is the solid angle subtended by the cold component. This condition,
which arises by assuming constant density and spherical symmetry for both the hot and cold
components, is imposed in order to overcome the degeneracy problem of the parameters
$N(\mathrm{H}_{2})_{\mathrm{hot}}$ and $\Omega_{\mathrm{d}}$, as was done for the SEDs fitting (see
Section~\ref{secSED}). We use as angular size of the cold cloud that determined from the 1.2~mm
emission \citepalias{Garay2007}.

\begin{deluxetable}{lcccc}
\tablecolumns{5}
\tabletypesize{\footnotesize}
\tablecaption{Parameters Derived from TIMMI2 Spectra \label{tableSpectra}}
\tablehead{
\colhead{Source} & \colhead{Flux [Ne II] \tablenotemark{a}} & \colhead{$T_{\mathrm{d}}$ \tablenotemark{b}} &
\colhead{$N(\mathrm{H}_2)_{\mathrm{cold}}$ \tablenotemark{b}} &
\colhead{$N(\mathrm{H}_2)_{\mathrm{1.2~mm}}$ \tablenotemark{c}}\\
\colhead{ } & \colhead{($10^{-15}$ W m$^{-2}$)} & \colhead{(K)} & \colhead{($10^{22}$ cm$^{-2}$)} & \colhead{($10^{22}$ cm$^{-2}$)}
}
\startdata
12383$-$6128  &\phn9.7 &  369  &\phn8.9 &\phn5.6\\
13291$-$6249  &  16.4  &  759  &  11.6  &\phn9.0\\
15520$-$5234  &\phn7.1 &  355  &  13.6  &  24.8\\
16128$-$5109  &  83.2  &  233  &\phn6.6 &\phn9.1\\
16458$-$4512  &  10.1  &  405  &  13.8  &  10.2\\
17008$-$4040~I&  13.9  &  339  &  17.1  &  21.4\\
17009$-$4042  &  31.1  &  458  &  13.8  &  32.2\\
17016$-$4124~I&\phn2.5 &\nodata\tablenotemark{d}& \nodata\tablenotemark{d}&   31.3\\
\enddata

\tablenotetext{a}{Uncertainties derived from the spectra are $\lesssim 2\%$ (only statistical
  error).}

\tablenotetext{b}{Best fit parameters; see Equation~(\ref{eq_flux_sil}). Uncertainties are $< 1\%$,
  but we consider them underestimated (see the text).}

\tablenotetext{c}{Computed from Equation~(\ref{eq_N_SIMBA}). Uncertainties are $\simeq 20\%$, only
  considering the error from the 1.2~mm flux.}

\tablenotetext{d}{~Low-quality spectrum.}

\end{deluxetable}

Columns 3 and 4 of Table \ref{tableSpectra} list the parameters $T_{\mathrm{d}}$ and
$N(\mathrm{H}_{2})_{\mathrm{cold}}$ obtained from the best fit to the spectrum extracted from the
central part of each source. The best fit for each source is shown in Figure~\ref{TIMMI2spectra} as
a dashed line. We assumed $R_{\mathrm{dg}} = 1/125$ \citep{Draine2003} and $\mu = 2.8$ (mean
molecular weight, adopting 10\% abundance of He with respect to H). The uncertainties derived from
the covariance matrix computed in the least-square minimization are $< 1\%$, but we suspect that
they are significantly underestimated, since the fitting only considers the statistical noise in the
spectra, which is also low.

To assess the robustness of the derived values of the parameters we investigated the effects
produced by changes in the assumptions. First, we used the \citeauthor{Weingartner&Draine2001} model
for the opacities with $R_V = 3.1$.  Since for the TIMMI2 spectral range, the $R_V = 5.5$ and $R_V =
3.1$ models do not present significant differences, we found that the fitting procedure is not
sensitive to such change, and the adjusted parameters only differ by $\simeq 5\%$ in
$T_{\mathrm{d}}$ and $\simeq 3\%$ in $N(\mathrm{H}_{2})_{\mathrm{cold}}$.

The average optical depths between 8 and 13~\micron\ inferred from the hot-component column
densities $N(\mathrm{H}_2)_\mathrm{hot}$ derived using the fiducial model, are in the range $ 0.01
\lesssim \langle \tau_\nu \rangle _\mathrm{8-13~\, \mu m} \lesssim 0.1 $, suggesting that the MIR
emission from hot dust is indeed optically thin in these regions.  Similar values for the optical
depths at 11.7~\micron\ were derived by \citet{DeBuizer2005a} from the MIR flux densities of their
sample of massive star-forming regions, which are mostly $\lesssim 0.3$.  Hence, we also
investigated the changes produced by assuming that the MIR emission is optically thin , i.e., using
\begin{equation}
F_\nu = A  \sigma_\nu B_\nu(T_{\mathrm{d}})
        e^{-\sigma_{\nu} N(\mathrm{H}_2)_{\mathrm{cold}}/2}\label{eq_flux_sil_thin}~~,
\end{equation}
where $A = N(\mathrm{H}_2)_\mathrm{hot} \Omega_\mathrm{d}$. Note that here the parameters
$N(\mathrm{H}_2)_\mathrm{hot}$ and $\Omega_\mathrm{d}$ cannot be fitted simultaneously, but this
allows us to avoid the constant-density condition (eq. [\ref{eq_ncte_sil}]). Under this assumption
we derived almost the same column densities $N(\mathrm{H}_{2})_{\mathrm{cold}}$ and temperatures
than those given by the fiducial model (differences $<1\%$).

Since in all eight sources the MIR emission is located at the peak of the millimeter core, the
fitted value of the cold dust cloud column density , $N(\mathrm{H}_{2})_{\mathrm{cold}}$, can be
compared with the column density, $N(\mathrm{H}_{2})_{\mathrm{1.2\, mm}}$, derived from the peak
value of the 1.2~mm flux density, $ S^{\mathrm{peak}}_{\mathrm{1.2\, mm}}$
\citepalias{Garay2007}. Assuming that the dust emission is optically thin at 1.2~mm, then
\begin{equation}
N(\mathrm{H}_{2})_{\mathrm{1.2\, mm}} = \frac{S^{\mathrm{peak}}_{\mathrm{1.2\, mm}}}
{\Omega_{\mathrm{B}}B_\mathrm{1.2\,mm}(T_{\mathrm{c}})\sigma_{\mathrm{1.2\, mm}}}~~,
\label{eq_N_SIMBA}
\end{equation}
where $\Omega_{\mathrm{B}}$ is the beam ($24''$) and $T_{\mathrm{c}}$ is the cold dust temperature
derived in Section~\ref{secSED}.  We use $\kappa_{\mathrm{1.2\,mm}} = 1$~cm$^2$~g$^{-1}$ from
\citet{Ossenkopf&Henning1994} , since the effect of coagulation and presence of dirty ice mantles on
dust grains in dense cold cores is likely to be important at this wavelength, where the
\citet{Weingartner&Draine2001} $R_V = 5.5$ model might not represent it properly (in particular, it
ignores ice mantles). The column densities calculated using this relation are given in Column 5 of
Table \ref{tableSpectra}.  The column densities derived from both methods are in good agreement,
within a factor of $\simeq 2$, particularly considering the uncertainties inherent to both methods
(e.g., in the opacities and dust-to-gas ratio), as well as the differences in angular resolution
between the TIMMI2 and 1.2~mm observations.

\section{Discussion}

\subsection{Spectral Energy Distributions}\label{secSED}

We constructed SEDs for all the 18 sources of our survey, collecting the flux densities at 1.2~mm
\citepalias[SIMBA data,][]{Garay2007}, at 12, 25, 60, and 100~\micron\ (\emph{IRAS} data), and at
8.3, 12.1, 14.7, and 21.3~\micron, computed from the images of the \emph{Midcourse Space Experiment
  (MSX)} Survey of the Galactic Plane \citep{Price2001}. We also included the IRAC (nonsaturated)
and TIMMI2 $F^{\mathrm{tot}}$ fluxes listed in Tables~\ref{photTIMMI2} and \ref{photIRAC}, summing
the flux of the two components for double sources. In most cases, a simple model consisting of
emission from three uniform gray bodies with different temperatures and sizes was able to fit the
SED.  Since the regions exhibit generally complex MIR structures and probably present temperature
gradients, this is indeed a coarse simplification. However, this model allows us to determine the
average dust temperatures representative of each wavelength range.

\begin{figure*}[!thp]

\centering
\includegraphics[height= 23cm]{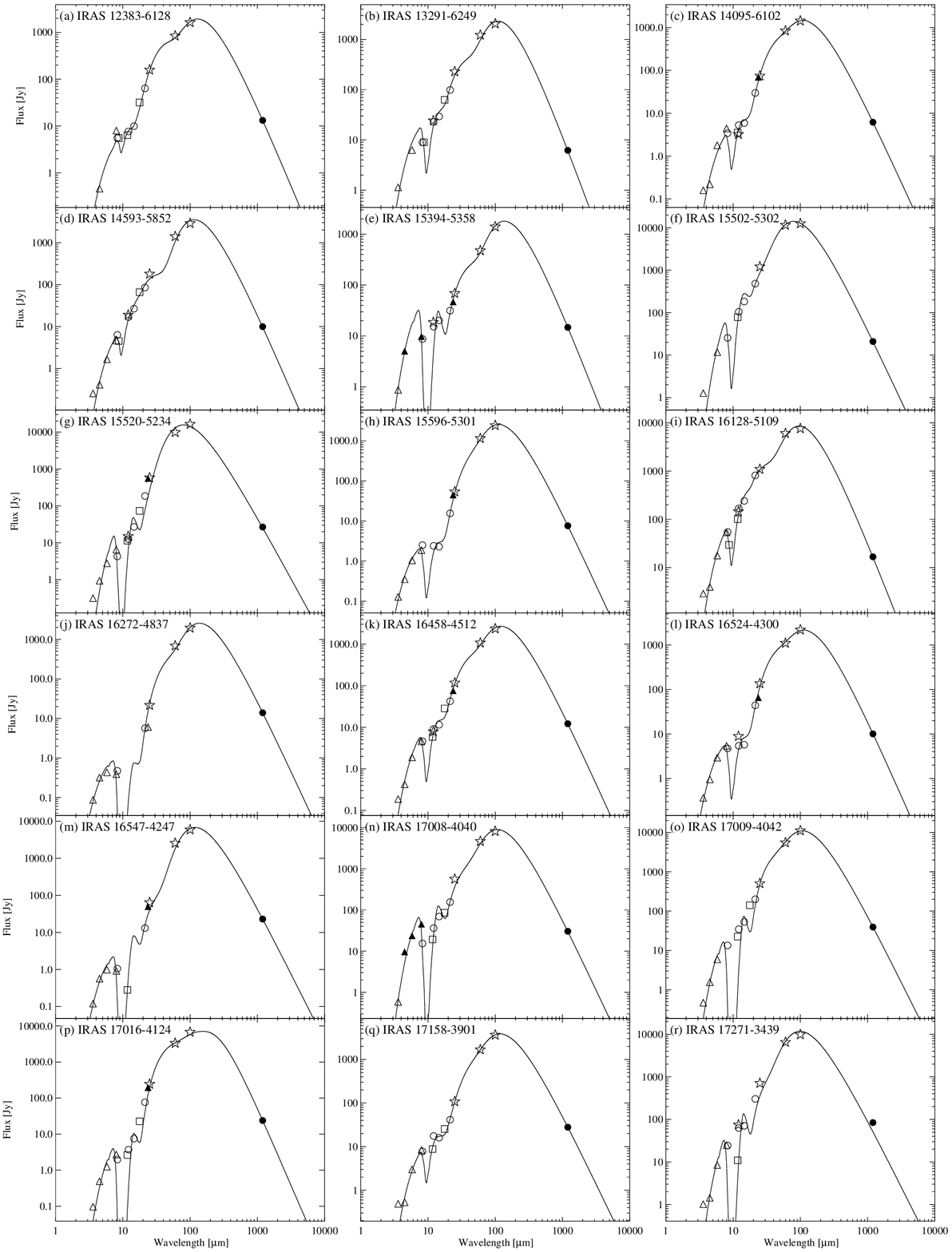}

\caption{SEDs and their corresponding fits as described in Section~\ref{secSED}. Flux symbols are a
  filled circle for 1.2~mm SIMBA, stars for \emph{IRAS}, empty circles for \emph{MSX}, squares for
  TIMMI2, and triangles for \emph{Spitzer}--IRAC/MIPS (filled if they are saturated fluxes
  satisfying Equation (\ref{eq_saturated})).}
\label{SEDfig}
\end{figure*}

The \emph{MSX} fluxes were computed just over the corresponding compact sources in the \emph{MSX}
images, in order to reduce the effect of the different resolution ($\simeq 18''$) with respect to
the IRAC data ($\simeq 1.8''$). Despite that difference, the SEDs show (Figure~\ref{SEDfig}) that
the \emph{MSX} fluxes are consistent with the IRAC and TIMMI2 fluxes (see, e.g., the good
correlation between \emph{MSX} 8.3~\micron\ fluxes and IRAC 8.0~\micron\ fluxes), indicating that
the major contribution to the ``compact'' \emph{MSX} flux arises from the brightest source(s) within
its beam. The \emph{IRAS} data, however, have a much coarser resolution, so they include emission
from a more extended region. Consequently, we often ignored the \emph{IRAS} 12~\micron\ flux for the
fitting, and at 1.2~mm we considered the whole flux to make it representative of the same region
that \emph{IRAS} accounts.  Conservative flux uncertainties of 20\% were taken for SIMBA,
\emph{IRAS}, \emph{MSX}, and IRAC data. For TIMMI2 data, we used the computed errors.

The model assumes the presence of three spherically symmetric components: a cold (c) extended
component, a warm (w) inner component, and a hot (h) compact component. We consider them as
radiating gray bodies and include the absorption of the radiation by the enclosing components. The
total flux density is then given by,
\begin{equation}
F_{\nu} = F_{\nu}^{\mathrm{h}} + F_{\nu}^{\mathrm{w}} + F_{\nu}^{\mathrm{c}}~~,
\label{eq_flux_SEDtot}
\end{equation}
where
\begin{eqnarray}\label{eq_flux_SEDcomp}
F_{\nu}^{\mathrm{h}} &=& \Omega_{\mathrm{h}}B_{\nu}(T_{\mathrm{h}})(1 - e^{-N_{\mathrm{h}}
 \sigma_{\nu}})e^{-(N_{\mathrm{c}} + N_{\mathrm{w}})\sigma_{\nu}/2}\label{eq_flux_SEDhot}\\
F_{\nu}^{\mathrm{w}} &=& \Omega_{\mathrm{w}}B_{\nu}(T_{\mathrm{w}})(1 - e^{-N_{\mathrm{w}}
 \sigma_{\nu}})e^{-N_{\mathrm{c}}\sigma_{\nu}/2}\label{eq_flux_SEDwarm}\\
F_{\nu}^{\mathrm{c}} &=& \Omega_{\mathrm{c}}B_{\nu}(T_{\mathrm{c}})(1 - e^{-(\nu/\nu_0)^{\beta}})~~,
\label{eq_flux_SEDcold}
\end{eqnarray}
where the superscripts c, w, and h denote quantities for the cold, warm, and hot components,
respectively; $\Omega$ is the solid angle; $T$ the dust temperature, and $N$ the H$_2$ column
density. The cross section per hydrogen molecule, $\sigma_{\nu}$, is given by Equation
(\ref{eq_sigma_nu}), where $R_{\mathrm{dg}} = 1/125$ and $\mu = 2.8$, and we adopted the
\citet{Weingartner&Draine2001} $R_V = 5.5$ model for dust opacities (see
Section~\ref{secTIMMI2spe}).  For the fit we considered as free parameters $\nu_0$, $\beta$,
$\Omega_{\mathrm{h}}$, $\Omega_{\mathrm{w}}$, $T_{\mathrm{h}}$, and $T_{\mathrm{w}}$. The parameters
of the cold component, $T_{\mathrm{c}}$, $\Omega_{\mathrm{c}}$, and $N_{\mathrm{c}}$ were taken as
fixed, and equal to the values listed in \citetalias[][]{Garay2007}.  Further, we adopted as upper
limits for $\Omega_{\mathrm{h}}$ and $\Omega_{\mathrm{w}}$, the values derived, respectively, from
the $N$-band observations (Table \ref{photTIMMI2}) and \emph{MSX} $E$-band observations. Because the
column densities for the hot and warm components are small, the expression $(1 - e^{-N
  \sigma_{\nu}})$ acts as a linear term in $N$, producing degeneracy between the parameters $N$ and
$\Omega$. To overcome this problem, we constrained the $N_{\mathrm{h}}$ and $N_{\mathrm{w}}$
parameters assuming constant density for all the temperature components, which implies
$N_{\mathrm{h,w}} = \sqrt{\Omega_{\mathrm{h,w}}/\Omega_{\mathrm{c}}}\,N_{\mathrm{c}}$ for the hot
and warm regions.  Given the simplicity of the SED model and the differences in the data used, a
more realistic density distribution is unnecessary.

\begin{deluxetable}{lrrrr}
\tablecolumns{5}
\tabletypesize{\footnotesize}
\tablecaption{Parameters Derived from SED Fitting \label{tableSED}}
\tablehead{
\colhead{Source} & \colhead{$T_{\mathrm{h}}$} & \colhead{$T_{\mathrm{w}}$} & 
\colhead{$\theta_{\mathrm{h}}$} & \colhead{$\theta_{\mathrm{w}}$}\\
\colhead{ } & \colhead{(K)} & \colhead{(K)} & \colhead{$('')$} & \colhead{$('')$}
}
\startdata
12383$-$6128  &  361  &   71  &  0.43  &  12.3\\
13291$-$6249  &  398  &   93  &  0.34  &   3.9\\
14095$-$6102  &  342  &   65  &  0.26  &   7.2\\
14593$-$5852  &  328  &  117  &  0.41  &   2.9\\
15394$-$5358  &  432  &   99  &  0.46  &   2.8\\
15502$-$5302  &  263  &  103  &  1.78  &   6.6\\
15520$-$5234  &  276  &  112  &  1.17  &   2.7\\
15596$-$5301  &  420  &   63  &  0.16  &   7.2\\
16128$-$5109  &  323  &  123  &  1.02  &   5.8\\
16272$-$4837  &  591  &   75  &  0.09  &   4.1\\
16458$-$4512  &  330  &   73  &  0.38  &   6.7\\
16524$-$4300  &  425  &   68  &  0.27  &   9.0\\
16547$-$4247  &  504  &  139  &  0.15  &   1.5\\
17008$-$4040  &  340  &   98  &  1.12  &   6.2\\
17009$-$4042  &  400\tablenotemark{a}  &  132  &  0.57  &   5.2\\
17016$-$4124  &  400\tablenotemark{a}  &   96  &  0.33  &   6.1\\
17158$-$3901  &  331  &   62  &  0.49  &  12.0\\
17271$-$3439  &  332  &  181  &  1.18  &   3.7\\
\enddata

\tablecomments{Best fit parameters, see Equations (\ref{eq_flux_SEDhot}) and
  (\ref{eq_flux_SEDwarm}).  Solid angles ($\Omega$) were converted to FHWMs ($\theta$) assuming
  Gaussian distributions. Statistical errors are about 5\%, 15\%, 12\%, and 30\% for
  $T_{\mathrm{h}}$, $T_{\mathrm{w}}$, $\theta_{\mathrm{h}}$, and $\theta_{\mathrm{w}}$,
  respectively.}

\tablenotetext{a}{Unstable fitting; reached upper limit of 400~K.}

\end{deluxetable}

Table \ref{tableSED} lists the parameters of the hot and warm components obtained from the best fit
to the SED using the model described above.  The FWHM angular sizes $\theta_{\mathrm{h}}$ and
$\theta_{\mathrm{w}}$ were computed from the fitted solid angles assuming Gaussian emission
distributions ($\Omega = 1.133\times\theta^2$). Covariance-matrix errors are about 5\%, 15\%, 12\%,
and 30\% for $T_{\mathrm{h}}$, $T_{\mathrm{w}}$, $\theta_{\mathrm{h}}$, and $\theta_{\mathrm{w}}$,
respectively, but they could be underestimated.  Most of the hot dust temperatures fall in the range
$\simeq 250-450$~K, whereas warm dust temperatures fall in the range $\simeq 60-150$~K.

Figure \ref{SEDfig} shows the SEDs and their corresponding fits for all sources. In most cases the
fit is excellent.  For the objects with the largest cold cloud column densities ($N_{\mathrm{c}} >
3\times10^{23}$ cm$^{-2}$), which gives rise to a very deep silicate absorption feature at
$\simeq$~10~\micron\ (e.g., IRAS~17271$-$3439, Figure~\ref{SEDfig}(r)), the inclusion of absorption
in the model in the MIR range became difficult. Because of this, particularly unstable fits were
obtained for IRAS~17009$-$4042 and 17016$-$4124 (Figures~\ref{SEDfig}(o) and (p),
respectively). Hence, for these objects we set an upper limit of 400~K for $T_{\mathrm{h}}$ in order
to obtain reasonable output parameters.  The SEDs of the sources IRAS~16272$-$4837 and 16547$-$4247
(Figures~\ref{SEDfig}(j) and (m)) exhibit a peculiar tendency in the IRAC fluxes, resulting in high
hot-dust temperatures (591~K and 504~K, respectively). In these sources, the IRAC 4.5~\micron\ flux
density is similar (only slightly lower) to the 5.8~\micron\ flux, which is higher than the
8.0~\micron\ flux. This trend is consistent with the high absorption and the green fuzzies found in
these regions.

For each source, the best-fit SED model was used as a validity check for the saturated IRAC/MIPS
fluxes, which were recovered by applying the \emph{imageworks} code (see Section~\ref{secGLIMPSE}),
but were not used by the SED fitting. We list in Table \ref{photIRAC}, and show in
Figure~\ref{SEDfig} as filled triangles, only those saturated fluxes that satisfy the condition
\begin{equation}
\left(\frac{F_{\lambda_0}^{\mathrm{sat}} - F_{\lambda_0}^{\mathrm{model}}}{\Delta F_{\lambda_0}}\right)^2 \le
\frac{1}{N} \sum_{i=1}^{N}  \left(\frac{F_{\lambda_{i}} - F_{\lambda_{i}}^{\mathrm{model}}}{\Delta
    F_{\lambda_{i}}}\right)^2 \label{eq_saturated}~~,
\end{equation}
where $F_{\lambda_0}^{\mathrm{sat}}$ is the recovered flux at the saturated wavelength $\lambda_0 =$
4.5, 5.8, 8.0, or 24~\micron, $F_{\lambda}^{\mathrm{model}}$ is the flux density given by the SED
model, $N$ is the number of data points being fitted, $\{F_{\lambda_{i}}\}$ are the observed fluxes,
and $\Delta F_{\lambda}$ represents the corresponding error (it was taken as 30\% for the saturated
flux). In other words, we request that $F_{\lambda_0}^{\mathrm{sat}}$ did not deviate from the SED
model more than the $\chi^2$-per-data point value of the fit. Using this criterion, it was found
that the \emph{imageworks} tool had about 50\% of success for embedded sources (e.g., seven out of
16 sources with saturated MIPS data).

\subsection{Warm Dust Regions and Association with Ionized Gas}\label{secMIRHII}

The high correlation between the spatial distribution of the MIR and radio emissions may suggest
that the MIR emission is due to the ionized gas.  However, extrapolating the observed radio
continuum flux of the optically thin \ion{H}{2} regions \citepalias{Garay2006} to the MIR, assuming
a $\nu^{-0.1}$ frequency dependence, we find that the free-free emission at MIR wavelengths is
negligible in comparison to the observed MIR fluxes (less than $\simeq 6\%$ at 8.0, 11.7, and
17.7~\micron).  We conclude that the MIR emission is tracing warm dust, located either
within the \ion{H}{2} region or in a thin shell around it.  \citet{Kraemer2003} already found
well-correlated MIR and radio distributions, and discarded the possibility that the emitting dust is
distributed in a thin shell, since in that case the MIR emission would show evidence of
limb-brightened morphology, which was not observed.  None of the regions of our sample exhibit
limb-brightened MIR emission.

We conclude that the MIR emission of the sources associated with radio-continuum emission arises
from warm dust mixed with the ionized gas inside the \ion{H}{2} region.  Although dust grains and
gas particles are likely to be dynamically coupled in \ion{H}{2} regions, the heating of the dust
grains is dominated by the UV radiation from the exciting source(s) and the ionized nebula, and not
by the collisional exchange of energy with the gas \citep[see, e.g.,][]{Spitzer1978}. This makes
possible the coexistence of a plasma with a kinetic temperature of $\simeq 7000$~K and solid
particles with a temperature of $\simeq 250-450$~K (Section~\ref{secSED}). These dust temperatures
are consistent with the models by \citet{Natta&Panagia1976} for dusty \ion{H}{2} regions.

\subsection{PAH Emission}\label{secPAH}

The $N$-band TIMMI2 images (and probably the $Q$-band images too) are sensitive mainly to the
emission of warm dust intimately associated with the ionized gas, or close to the exciting sources
when there is no radio emission. They do not trace the presence of PAH, as indicated by the
$8-13$~\micron\ spectroscopy toward seven of the eight sources of our sample (all but
IRAS~12383$-$6128), where the emission bands in this wavelength range (the more prominent at 8.6 and
11.3~\micron) are not seen.

Within the \ion{H}{2} region or very close to a massive protostar, hard UV radiation destroys these
molecules. For example, \citet{Povich2007} calculated the photodestruction rate of PAHs by
extreme-UV (EUV) photons and found that this destruction mechanism accounts for the absence of PAHs
in the M17 \ion{H}{2} region. Because this region is one of the most energetic in the Galaxy, we
expect that for the UC~sources of our sample, the PAH destruction edge is located much closer to the
exciting star(s). IRAS 12383$-$6128, which is the only one that presents PAH emission bands at 8.6
and 11.3~\micron, has probably a softer UV spectrum than the others.

It is possible that a part of the outer diffuse emission observed in the TIMMI2 images is due to
PAHs, but since the slit was centered in the bright compact MIR sources, the PAH radiation was not
strong enough to be detected. This explanation is supported by other MIR-spectroscopic observations
found in the literature, with larger apertures, toward some of our sources, in which PAH emission is
detected. In particular, the \emph{IRAS} LSR spectrum (aperture of $6'\times5'$) of
IRAS~13291$-$6249 \citep{Jourdain1990} and the ISO spectrum (aperture of $14''\times20''$ in the
wavelength range of TIMMI2) of IRAS~16128$-$5109 \citep{Peeters2002} reveal the presence of PAH
emission bands within the wavelength range of the TIMMI2 spectra (at 8.6 and 11.3~\micron). For
IRAS~16128$-$5109, the PAH bands at 3.3, 6.2, 7.7, and 12.7~\micron\ are also present
\citep{Peeters2002}. PAH emission was also detected toward other sources of our sample which were
not observed spectroscopically by us: IRAS~14593$-$5852 \citep[IRAS LSR,][]{Zavagno1992},
IRAS~15502$-$5302 \citep[ISO,][]{Peeters2002}, IRAS~15596$-$5301, and IRAS~17271$-$3439 \citep[IRAS
LSR,][]{Jourdain1990}.

The latter result strongly suggests that the extended diffuse emission seen in most regions by the
\emph{Spitzer}--IRAC images beyond the radio continuum contours include an important contribution of
PAH emission bands. The [3.6], [5.8], and [8.0] IRAC filters cover the PAH features at 3.3, 6.2, and
7.7~\micron, respectively. The [8.0] filter also covers the PAH band at 8.6~\micron. This extended
diffuse emission is, however, more clearly seen in the [5.8] and [8.0] IRAC images than in the [3.6]
images. A possible explanation \citep{Povich2007} is that the 6.2~\micron\ feature might have a
greater fractional contribution to the [5.8] filter than the 3.3~\micron\ band to the [3.6] filter
(and the same for the [8.0] and [3.6] filters), or simply that the 3.3~\micron\ band is
intrinsically fainter. As for the TIMMI2 images, the \emph{Spitzer} MIR emission at the location of
the UC~\ion{H}{2} regions is more likely to correspond to warm dust continuum radiation with little
or any contribution of PAH features.

\subsection{High-Mass Protostellar Objects}\label{secHMPO}

The high angular resolution MIR observations presented here revealed the presence of bright compact
MIR sources in the vicinity of, but not coincident with, UC \ion{H}{2} regions, which are likely to
signpost HMPOs. We identify a total of five new HMPO candidates within the regions observed with
TIMMI2: IRAS 14593$-$5852~II (only detected at 17.7~\micron, see Figure~\ref{14593}), 17008$-$4040~I
and II (Figure~\ref{17008}), 17016$-$4124~II (Figure~\ref{TIMMI2images2}(d)), and the compact
component of IRAS 17158$-$3901 (Figure~\ref{TIMMI2images2}(e)).

In order to estimate the spectral types of these HMPO candidates, we derived MIR luminosities from
the measured $N$- and $Q$-band fluxes, following a method based on the work by
\citet{DeBuizer2000}. This technique provides good lower limits to the true bolometric luminosities
of the stellar sources \citep[see][]{DeBuizer2002b}.  We note that estimates made using the
\emph{IRAS} fluxes are inappropriate since they include the contribution from the exciting star(s)
of the nearby UC~\ion{H}{2} regions, as well as from other IR sources within the large \emph{IRAS}
beam. First, a dust color temperature is computed assuming that the MIR emission from warm dust is
optically thin, as we suggested in Section~\ref{secTIMMI2spe}. Our SED fits are also consistent with
optically thin emission from the hot-dust component (using the derived $N_{\mathrm{h}}$, the optical
depths at 10~\micron\ are in the range $\simeq 0.01-0.3$). We adopt a standard expression for the
flux density (analog to Equation (\ref{eq_flux_sil_thin})):
\begin{equation}
F_{\nu} = A \sigma_{\nu} B_{\nu}(T_{\mathrm{color}}) e^{-N_{\mathrm{c}}\sigma_{\nu}/2}~~.
\label{eq_flux_Tcolor}
\end{equation}
Here, the factor $A = N_\mathrm{h} \Omega$ is constant, $\sigma_\nu$ is given by
Equation~(\ref{eq_sigma_nu}), and $N_{\mathrm{c}}$ by Equation~(\ref{eq_N_SIMBA}), but truncated to
the maximum column density derived from TIMMI2 spectra if it is higher than that value ($17.1\times
10^{22}$~cm$^{-2}$, see Table~\ref{tableSpectra}), in order to moderate the effect of dust
absorption (see below). The color temperature $T_{\mathrm{color}}$ is derived from the ratio of the
integrated fluxes at the $N$- and $Q$-band (lower limit $N$-band flux for IRAS 14593$-$5852~II):
\begin{equation}
\frac{e^{hc/(k\lambda_2 T_{\mathrm{color}})} - 1}{e^{hc/(k\lambda_1 T_{\mathrm{color}})} - 1}
 = \frac{F_{\nu_1}}{F_{\nu_2}} \left(\frac{\lambda_1}{\lambda_2}\right)^3
  \frac{\kappa_{\nu_2}}{\kappa_{\nu_1}} e^{-N_{\mathrm{c}}(\sigma_{\nu_2} - \sigma_{\nu_1})/2}~~.
\label{eq_Tcolor}
\end{equation}
The MIR luminosity is then obtained integrating an expression of the form $F_{\nu} = A \sigma_{\nu}
B_{\nu}(T_{\mathrm{color}})$ from 1 to 1000~\micron\, using the kinematical distance and the factor
$A$ to scale the flux density to the observed ones. Columns 2 and 3 of Table~\ref{tabHMPO} give the
computed color temperatures~$T_\mathrm{color}$ and MIR luminosities~$L_\mathrm{MIR}$,
respectively. We find that the derived spectral types, using the calibration of \citet{Panagia1973}
and the MIR luminosities, are earlier than B3 for four of our HMPO candidates (Column 4 of
Table~\ref{tabHMPO}).

\begin{deluxetable}{lcrc}
\tablecolumns{4}
\tabletypesize{\footnotesize}
\tablecaption{Characteristics of the New HMPO Candidates \label{tabHMPO}}
\tablehead{
\colhead{Source} & \colhead{$T_\mathrm{color}$} & \colhead{$L_\mathrm{MIR}$} & \colhead{Spectral Type} \\
\colhead{ } & \colhead{(K)} & \colhead{($L_\sun$)} & \colhead{ }
}
\startdata
14593$-$5852~II\tablenotemark{a}  &  142  &  54500  &  O8.5  \\
17008$-$4040~I   &  219  &  40900  &  O9.5  \\
17008$-$4040~II  &  199  &   2490  &  B2    \\
17016$-$4124~II  &  170  &   6120  &  B1    \\
17158$-$3901     &  169  &    542  &  \nodata\tablenotemark{b} \\
\enddata

\tablecomments{Parameters $T_\mathrm{color}$ and $L_\mathrm{MIR}$ were computed from the observed
  $N$- and $Q$-band fluxes. The spectral types were derived from $L_\mathrm{MIR}$ (and thus are
  lower limits) and the calibration of \citet{Panagia1973}.}

\tablenotetext{a}{~Lower limit $N$-band flux used.} 

\tablenotetext{b}{Spectral type later than B3.}

\end{deluxetable}

The spectral types derived by us are earlier than the typical spectral types obtained by
\citet{DeBuizer2000} using a similar method. A substantial difference between our technique and
theirs is that we have considered the absorption of a cold cloud to model the observed fluxes, and
corrected by extinction when computing the MIR luminosities. We found, however, that the employed
method is quite sensitive to assumed column density $N_c$ of the absorbing cloud, and therefore, in
order to obtain strict lower limits for the spectral types, we recomputed the MIR luminosities under
the assumption that there is no extinction ($N_c = 0$). This results now in two sources earlier than
B3, namely, B0 for IRAS 14593$-$5852~II and B2 for 17008$-$4040~I, which thus might be considered as
genuine HMPOs.

Figures~\ref{14593} and \ref{17008} present the whole set of MIR images for the regions containing
the two bona fide O-type high-mass protostar candidates: IRAS 14593$-$5852~II and
17008$-$4040~I. The \emph{Spitzer}--IRAC images toward IRAS~14593$-$5852 (Figure~\ref{14593}(a))
show only diffuse emission toward the HMPO, bright in red (8.0~\micron) and thus probably
corresponding to the environment's PAH emission, which dominates the large-scale structure.
IRAS~14593$-$5852~II is then likely to be a deeply embedded object. IRAS~17008$-$4040~I is bright in
all available MIR images, even saturated in the IRAC filters (Figure~\ref{17008}(a)). The
\ion{H}{2} region is displaced by $\simeq$~30$''$ to the west of the TIMMI2 sources and associated
with extended diffuse emission at 8.0~\micron. In addition to its high MIR luminosity and the lack
of a detectable \ion{H}{2} region (as indicated by the lack of both radio continuum emission and of
the [\ion{Ne}{2}] line in its $N$-band spectrum), there is additional evidence supporting the nature
of IRAS~17008$-$4040~I as a high-mass protostar: (i) the \emph{Spitzer}--IRAC image shows the
presence of extended filamentary green fuzzies (see Section~\ref{secGLIMPSE}) to its northeast,
suggesting outflowing activity; (ii) it is located at the center of a massive and dense molecular
core \citepalias[peak of 1.2~mm emission,][]{Garay2007}, as predicted by all the massive star
formation theories; and (iii) it is associated with different types of maser emission
(Figure~\ref{17008}(b)), all of which are known to be associated with high-mass star-forming
regions: OH masers at 1665/1667~MHz \citep[e.g.,][]{Forster&Caswell1989, Caswell1998} and 6.7~GHz
methanol masers \citep[e.g.,][]{Walsh1998} just coincident with the position of IRAS~17008$-$4040~I,
and water vapor maser emission at 22~GHz located $\simeq 4''$ to its southeast
\citep{Forster&Caswell1989}. In particular, 6.7~GHz methanol masers are only detected toward
high-mass star-forming regions \citep{Minier2003}.

Finally, although the MIR luminosities given in Table \ref{tabHMPO} should only be considered as
rough estimates, it is interesting to know where the HMPO candidates are located in the
$L_{\mathrm{bol}}-M_{\mathrm{env}}$ (bolometric luminosity - envelope mass) diagram presented in the
work of \citet[][their Figure~9]{Molinari2008}. They successfully used it to follow the
pre-main-sequence evolution of massive young stellar objects, by classifying a numerous sample of
objects according to SED modeling and then comparing with the predictions of the turbulent core
model \citep{McKnee&Tan2003}. We can do such exercise only for IRAS~17008$-$4040~I, since it is the
only HMPO candidate that is far enough from nearby UC~\ion{H}{2} regions and thus the mass of the
associated 1.2~mm core can be considered as its envelope mass (neglecting also the secondary object
IRAS~17008$-$4040~II, which is much less luminous). Then, assuming $L_{\mathrm{bol}} = 40900~L_\sun$
(MIR luminosity) and $M_{\mathrm{env}} = 1200~M_\sun$ \citepalias{Garay2007}, IRAS~17008$-$4040~I
lies in a zone dominated by MM-P objects, below that occupied by IR-P objects in the notation of
\citet{Molinari2008}. They suggested that the IR-P stage may correspond to the arrival of the
protostar on the zero-age main sequence (ZAMS), and the MM-P objects represent an earlier stage. The
location of IRAS~17008$-$4040~I in the $L_{\mathrm{bol}}-M_{\mathrm{env}}$ plot would then
categorize it as a pre-ZAMS object.

\section{Summary and Conclusions}

We present MIR $N$- and $Q$-band imaging, made with the TIMMI2 camera, toward 14 luminous
\emph{IRAS} point sources with colors of UC~\ion{H}{2} regions, thought to be massive star-forming
regions in early stages of evolution.  We also present $N$-band spectroscopic observations toward
eight of these regions. Images from the \emph{Spitzer} legacy programs GLIMPSE and MIPSGAL for the
18 sources of our multiwavelength survey complement our data set. The main results and conclusions
presented in this paper are summarized as follows.

The morphology of the TIMMI2 emission toward most regions (10) is simple, exhibiting either one
(seven cases) or two (three cases) compact components each surrounded by extended emission.  The
compact components have physical sizes in the range $0.008-0.18$~pc. The \emph{Spitzer}--GLIMPSE
images of these regions show even more extended diffuse emission than that seen by TIMMI2, most
likely due to emission from PAH bands.

We find that the MIR emission traced by TIMMI2 corresponds to optically thin radiation from hot
dust. The hot-dust temperatures, derived from fitting the SEDs, are typically $\simeq 250-450$~K.

We find that the spatial distribution of the MIR emission and radio continuum emission (Paper I) are
highly correlated, suggesting that the ionized gas (radio source) and hot dust (MIR source) are
intimately associated. We conclude that the emitting dust is located inside the \ion{H}{2} region,
mixed with the ionized gas.  The $N$-band spectra toward seven of the eight sources exhibit bright
[\ion{Ne}{2}] line emission, consistent with the association of these sources with ionized gas.

The $N$-band spectra show, for all eight observed sources, the presence of a broad and deep silicate
absorption band at 9.7~\micron, confirming their nature of deeply embedded objects within massive
dense cores of cold dust and molecular gas.  The H$_2$ column densities derived from this absorption
feature are in the range $(7-17)\times 10^{22}$~cm$^{-2}$, and are in good agreement (within a
factor of $\simeq 2$) with those estimated from 1.2~mm data \citepalias{Garay2007}. Only toward one
source, IRAS 12383$-$6128, we detected the presence of PAH emission at 8.6 and 11.3~\micron.  The
lack of PAH emission in the spectra, with a small aperture and centered in the bright compact
object, from the remaining sources is likely due to the presence of a hard radiation field which
destroys these molecules in their neighborhood.

We discovered five bright compact MIR sources which are not associated with radio continuum
emission, and are thus prime candidates for hosting young massive protostars.  In particular,
objects IRAS 14593$-$5852~II (\emph{only} detected at 17.7~\micron) and 17008$-$4040~I are likely to
be genuine O-type protostellar objects.

\acknowledgements

E.F.E.M., D.M., G.G., K.J.B., and J.E.P. gratefully acknowledge support from the Chilean Centro de
Astrof\'\i sica FONDAP No. 15010003 and Proyecto BASAL PFB-06.


\bibliography{paperIII}

\end{document}